\documentclass[twocolumn,aps,prd]{revtex4}
 \usepackage{graphicx,amssymb}
 \usepackage[utf8]{inputenc} 
 
\usepackage[urlcolor=blue,colorlinks,breaklinks=true]{hyperref}
\PassOptionsToPackage{obeyspaces,spaces,hyphens}{url}
 
 \usepackage{amsmath}

\begin{document}

 	\def\half{{1\over2}}
 	\def\shalf{\textstyle{{1\over2}}}
 	
 	\newcommand\lsim{\mathrel{\rlap{\lower4pt\hbox{\hskip1pt$\sim$}}
 			\raise1pt\hbox{$<$}}}
 	\newcommand\gsim{\mathrel{\rlap{\lower4pt\hbox{\hskip1pt$\sim$}}
 			\raise1pt\hbox{$>$}}}

\newcommand{\be}{\begin{equation}}
\newcommand{\ee}{\end{equation}}
\newcommand{\bq}{\begin{eqnarray}}
\newcommand{\eq}{\end{eqnarray}}

\title{Sound speed as a source of the gravitational field in modified gravity}
 	 	
\author{P.P. Avelino}
\email[Electronic address: ]{pedro.avelino@astro.up.pt}
\affiliation{Departamento de F\'{\i}sica e Astronomia, Faculdade de Ci\^encias, Universidade do Porto, Rua do Campo Alegre 687, PT4169-007 Porto, Portugal}
\affiliation{Instituto de Astrof\'{\i}sica e Ci\^encias do Espa{\c c}o, Universidade do Porto, CAUP, Rua das Estrelas, PT4150-762 Porto, Portugal}
\affiliation{Université Côte d’Azur, Observatoire de la Côte d’Azur, CNRS, Laboratoire Lagrange, France}

\date{\today}
\begin{abstract}
	
In the context of $f(R,T)$ gravity and other modified theories of gravity, the knowledge of the first order variation of the trace $T$ of the energy-momentum tensor with respect to the metric is essential for an accurate characterization of the gravitational field. In this paper, by considering a paradigmatic example of a perfect fluid whose dynamics is described by a pure k-essence matter Lagrangian in $f(R,T)=R+\mathcal F(T)$ gravity, we show that the first order variation of the trace of the energy-momentum tensor cannot in general be determined from the proper density, proper pressure and 4-velocity of the fluid alone, and that the sound speed of the fluid can directly influence the dynamics of gravity. We also confirm that the second variation of the matter Lagrangian with respect to the metric should not in general be neglected. These results can be  particularly relevant for cosmological studies of $f(R,T)$ gravity in which some of the material content of the Universe is modeled as a perfect fluid.
	
\end{abstract}

\maketitle
 	
\section{Introduction}
\label{sec:intr}

The evidence for the current acceleration of the Universe is overwhelming \cite{Planck:2018vyg,eBOSS:2020yzd,Brout:2022vxf,DES:2024tys,DESI:2024mwx}. There are also strong reasons to believe that a period of acceleration in the early Universe, might be the solution to some of the most profound cosmological conundrums (see, for example, \cite{Planck:2018jri} and references there in). In general relativity a period of acceleration, no matter how short, requires the Universe to be dominated by a dark energy component violating the strong energy condition. Although dark energy may be the real origin for the early and late time acceleration of the expansion of the Universe, the true cause may be more profound and require an alternative theory of gravity \cite{Clifton:2011jh, Berti:2015itd, Nojiri:2017ncd}.

The exploration of extensions of general relativity, aimed at providing more natural explanations for the early and late time dynamics of the Universe, is an extremely active area of research \cite{Nojiri:2010wj,Olmo:2011uz,Clifton:2011jh,Capozziello:2011et,Ludwig:2015hta,Berti:2015itd,Nojiri:2017ncd}. Within this realm, significant attention is directed towards broad categories of modified theories of gravity that consider the possibility of a nonminimal coupling between geometry and matter \cite{Harko:2010mv,Harko:2011kv,Haghani:2013oma, Harko:2018gxr, Avelino:2018rsb,Azevedo:2018nvi,Haghani:2023uad,Goncalves:2023umv}. In some of these, such as in the case of $f(R,T)$ gravity, the dynamics of the gravitational and matter fields may depend on the first variation of the trace of the energy-momentum tensor with respect to the metric \cite{Harko:2011kv,Haghani:2023uad}.

The material content of the Universe is often described as a collection of fluids, which, as will happen in the present article, are frequently assumed to be perfect  \cite{Ferreira:2020fma}. Although in the context of general relativity the sound speed of these fluids does not explicitly appear in the Einstein equations, we will show that this is not generally the case in the context theories of gravity in which the dynamics of the gravitational and matter fields depends on the first variation of the trace of the energy-momentum tensor with respect to the metric. In this paper $f(R,T)$ gravity will be considered as a prime example of such theories. 

In \cite{Haghani:2023uad}  it has been claimed that the second variation of the matter Lagrangian of a perfect fluid with respect to the metric tensor cannot generally neglected in the context of $f(R,T)$ gravity. The authors also argue that the first variation of the matter energy–momentum tensor with respect to the metric tensor can be expressed in terms of the pressure, the energy–momentum tensor itself, and the matter fluid 4-velocity. In the present paper we will revisit these issues, considering a perfect fluid whose dynamics is described by a pure k-essence matter Lagrangian in $f(R,T)=R+\mathcal F(T)$ gravity. This will turn out to be a well controlled model since it will be shown to be equivalent to a pure k-essence model of an isentropic, irrotational perfect fluid with conserved particle number and constant entropy per particle in general relativity. We shall assess the contribution of the second variation of the matter Lagrangian of a perfect fluid with respect to the metric tensor, and confirm that it cannot generally be neglected in the context of $f(R,T)$ gravity. Furthermore, we will investigate the potential impact of the sound speed on the dynamics of the gravitational and matter fields. We shall see that, although in general relativity the sound speed does not explicitly appear as a source of gravity, this might not be the case in the context of $f(R,T)$ gravity and other theories of gravity.

Throughout this paper, we will work in units where $c=16\pi G = \hbar=1$ with $c$, $\hbar$, and $G$ being, respectively, the speed of light in vacuum, the reduced Planck constant, and Newton's gravitational constant. We also adopt the metric signature $(-,+,+,+)$. The Einstein summation convention will be used when a Greek index appears twice in a single term, once in an upper (superscript) and once in a lower (subscript) position.

\section{$f(R,T)$ gravity}

Consider the action
\begin{equation}
\label{eq:action}
S=\int d^4 x \sqrt{-g} \left[ f(R,T) + \mathcal{L}_{\rm m}\right]\,,
\end{equation}
where $g$ is the determinant of the metric $g_{\mu\nu}$, $\mathcal{L}_{\rm m}$ is the Lagrangian of the matter fields, and $f(R,T)$ is a generic function of the Ricci scalar $R$ and of the trace of the energy-momentum tensor $T$. The corresponding equations of motion for the gravitational field are given by \cite{Harko:2011kv}
\be
2(R_{\mu\nu} - \Delta_{\mu\nu})f_{,R} - g_{\mu\nu}R={\mathfrak T}_{\mu\nu}\, ,
\ee
where $R_{\mu\nu}$ is the Ricci tensor, $R=g^{\alpha \beta} R_{\alpha\beta}$, $\Delta_{\mu \nu} \equiv \nabla_\mu \nabla_\nu - g_{\mu \nu} \Box$, $\Box \equiv \nabla^\mu \nabla_\mu$,  a comma denotes a partial derivative, and 
\be
{\mathfrak T}_{\mu\nu}=  T_{\mu\nu}+(f-R)g_{\mu\nu}-2f_{,T}(T_{\mu\nu}+{\mathbb  T}_{\mu\nu}) \,.
\ee
Here, 
\be
{\mathbb{T}}_{\mu \nu} = g^{\alpha \beta}\frac{\delta T_{\alpha \beta} }{\delta g^{\mu \nu} }\,,
\ee
and the components of the energy-momentum tensor,
\be
T_{\mu\nu} = -{2\over \sqrt{-g}}{\delta(\sqrt{-g}\mathcal{L}_{\rm m})\over \delta g^{\mu\nu}}= g_{\mu \nu} \mathcal{L}_{\rm m} -2 \frac{\delta\mathcal{L}_{\rm m}}{ \delta g^{\mu\nu}}\,, \label{EMTC}
\ee
are related by
\be
\frac{\delta T}{\delta g^{\mu \nu}}=T_{\mu\nu}+\mathbb{T}_{\mu \nu}\,,\label{delT}
\ee
where $T=g^{\alpha \beta} T_{\alpha \beta}$. 
Using Eq. \eqref{EMTC}, it can also be shown that
\be
\frac{\delta T}{\delta g^{\mu \nu}}=-T_{\mu \nu} +g_{\mu \nu} \mathcal{L}_{\rm m} - 2 g^{\alpha \beta} \frac{\delta^2 \mathcal{L}_{\rm m}}{\delta g^{\mu \nu} \delta g^{\alpha \beta} }\,, \label{D2L1}
\ee
or, equivalently, that
\be
\mathbb{T}_{\mu \nu}=-2T_{\mu \nu} +g_{\mu \nu} \mathcal{L}_{\rm m} - 2 g^{\alpha \beta} \frac{\delta^2 \mathcal{L}_{\rm m}}{\delta g^{\mu \nu} \delta g^{\alpha \beta} }\,. \label{D2L2}
\ee

\subsection{$R+\mathcal F(T)$ gravity}

If $f_{,R}=1$, then $f(R,T)=R+\mathcal F(T)$, where $\mathcal F(T)$ is a generic function of $T$. In this case, the Einstein equations,
\be
G_{\mu\nu}=R_{\mu\nu} -\frac12 g_{\mu\nu}R=\frac12 {\mathfrak T}_{\mu\nu} \, ,
\ee
are satisfied except for the replacement $T_{\mu \nu} \to {\mathfrak{T}}_{\mu \nu}$. Here, 
\be
{\mathfrak T}_{\mu\nu}=  T_{\mu\nu}+\mathcal F g_{\mu\nu}-2{\mathcal F}_{,T}(T_{\mu\nu}+{\mathbb  T}_{\mu\nu}) \label{TGR}
\ee
is covariantly conserved, so that
\be
\nabla^\mu {\mathfrak T}_{\mu\nu}=0\,.
\ee
In $R+\mathcal F(T)$ gravity
\be
{\mathfrak T_{\mu\nu}} = -{2\over \sqrt{-g}}{\delta\left[\sqrt{-g}(\mathcal{L}_{\rm m}+\mathcal F)\right]\over \delta g^{\mu\nu}}\,, \label{TN}
\ee
thus highlighting the fact that this model of gravity is totally equivalent to general relativity with the modified matter Lagrangian $\mathfrak L_{\rm m}=\mathcal{L}_{\rm m}+\mathcal F$. The equivalence with general relativity makes this model specially suited for investigating generic features of $f(R,T)$ gravity in a well controlled manner. We will then consider this model of gravity in the remainder of this paper.

\section{The role of the sound speed}

Here, we start by considering a scalar field $\phi$ described by a pure k-essence Lagrangian $\mathcal{L}_{\rm m}\left(X\right)$, where
\be
X=-\frac12 \nabla_\mu \phi \nabla^\mu \phi \,.
\ee
If $X>0$, the components of the associated energy-momentum tensor, given by
\be
T_{\mu \nu}=\mathcal L_{{\rm m},X} \nabla_\mu \phi \nabla_\nu \phi + {\mathcal L}_{\rm m} g_{\mu \nu}\,, \label{EM1}
\ee
may be written in a perfect fluid form 
\be
T_{\mu\nu}=\left(\rho+p\right)u^{\mu}u^{\nu}+pg^{\mu\nu}\,.\label{PF}
\ee
In Eq. \eqref{PF}, $\rho=2X \mathcal L_{{\rm m},X} - \mathcal L_{\rm m}$ and $p=\mathcal L_{\rm m}$ represent, respectively, the proper energy density and the proper pressure of the fluid, and $u_\mu = -\nabla_\mu \phi /\sqrt{2X}$ are the components of its 4-velocity (satisfying $u^\mu  u_\mu=-1$). The trace of the energy-momentum tensor is a function of $X$ alone, being equal to
\be
T= -\rho+3p=-2 X\mathcal L_{{\rm m},X}  + 4 {\mathcal L}_{\rm m} \,.
\ee

Here, we will start by writing the last term in Eqs. \eqref{D2L1} and \eqref{D2L2} as a function of the physical variables of the fluid. Taking into account that
\be
\frac{\delta \mathcal{L}_{\rm m}}{\delta g^{\mu \nu} }= \mathcal{L}_{\rm m,X}\frac{\delta X}{\delta g^{\mu \nu} } = -\frac12 \mathcal{L}_{\rm m,X} \nabla_\mu \phi \nabla_\nu \phi\,,
\ee
it can be shown that
\bq
- 2 g^{\alpha \beta} \frac{\delta^2 \mathcal{L}_{\rm m}}{\delta g^{\mu \nu} \delta g^{\alpha \beta} } &=& X \mathcal L_{{\rm m},XX} \nabla_\mu \phi \nabla_\nu \phi  \nonumber \\
&=&  2X^2 \mathcal L_{{\rm m},XX} u_\mu  u_\nu \nonumber\\
&=&\frac{\rho+p}{2}\left(\frac{1-c_s^2}{c_s^2}\right)  u_\mu u_\nu\,,\label{D2L3}
\eq
where the sound speed squared is given by
\be
c_s^2 = \frac{dp}{d\rho}=\frac{p_{,X}}{\rho_{,X}} = \frac{\mathcal{L}_{{\rm m},X}}{2X\mathcal{L}_{{\rm m},XX}+\mathcal{L}_{{\rm m},X}}\,.
\label{key}
\ee
Hence, we confirm the claim made in \cite{Haghani:2023uad} that the second variation of the matter Lagrangian of a perfect fluid with respect to the metric tensor components cannot generally neglected in the context of $f(R,T)$ gravity. However, it provides a contribution which cannot be inferred from the knowledge of the the proper density, proper pressure and 4-velocity alone, since it also depends on the sound speed of the fluid. Notice that this term is never zero, except if $c_s^2=1$. Also notice that all the calculations leading to the result given in Eq. \eqref{D2L3} were made off shell (this is in fact required in order to obtain the correct result). 

Using Eqs. \eqref{D2L2}, \eqref{EM1}, \eqref{PF} and \eqref{D2L3}, one finds that 
\be
\mathbb{T}_{\mu \nu}=\frac{\rho+p}{2} \left( \frac{1-5c_s^2}{c_s^2}\right)u_\mu u_\nu - p g_{\mu \nu}\,, \label{TT}
\ee
or, equivalently, that
\be
\frac{\delta T}{\delta g^{\mu \nu}}=\frac{\rho+p}{2} \left( \frac{1-3c_s^2}{c_s^2}\right)u_\mu u_\nu \,.
\ee

On the other hand, Eqs. \eqref{TGR}, \eqref{PF}, and \eqref{TT} imply that
\bq
{\mathfrak T}_{\mu \nu}&=&\left(\tilde \rho+\tilde p\right) u_\mu u_\nu + \tilde p  g_{\mu \nu}\nonumber\\
 &=&\left(\left(\rho+p\right)\left[1+\mathcal F_{,T}\left(\frac{3c_s^2-1}{c_s^2}\right)\right]\right) u_\mu u_\nu \nonumber \\
&+&( p+\mathcal F ) g_{\mu \nu}\,,
\eq
where
\bq
\tilde \rho&=&\left(\rho+p\right)\left[1+\mathcal F_{,T}\left(\frac{3c_s^2-1}{c_s^2}\right)\right] -(p +\mathcal F) \,, \\
\tilde p &=& p+\mathcal F \,.
\eq
Again notice that our model is equivalent to general relativity with the modified matter Lagrangian $\mathfrak L_{\rm m}=\mathcal{L}_{\rm m}+\mathcal F$, in which case, the ${\mathfrak T}_{\mu \nu}$ would represent the components of the corresponding energy-momentum tensor. 

Consider the identifications
\bq
\tilde p&=& \mathfrak{L}_{\rm m}\,, \label{ptilde}\\
\tilde \rho &=& 2X\mathfrak{L}_{{\rm m},X}-\mathfrak{L}_{\rm m}\,,\label{rhotilde}\\
u_\mu&=&-\nabla_\mu \phi/{\sqrt{2X}}\,,\\
\tilde \mu&=&\sqrt{2X} \,, \\
\tilde n&=& \sqrt{2X}{\mathfrak L}_{{\rm m},X}\,, \label{ntilde}
\eq
where $\tilde \rho$, $\tilde p$, $u_\mu$, $\tilde n=d \tilde p/d \tilde \mu$ and $\tilde \mu=d\tilde \rho/d \tilde n$ define, respectively, the  proper energy density, proper pressure, 4-velocity, proper particle number density and chemical potential of the perfect fluid described by the matter Lagrangian $\mathfrak L_{\rm m}(X)$. The equation of motion of the scalar field
\be
\nabla_{\mu}\left(\mathfrak{L}_{{\rm m},X}\nabla^{\mu}\phi\right) = \nabla_{\mu}\left(\tilde nu^\mu\right)= 0\,,
\ee
ensures particle number conservation. 
On the other hand, Eqs. \eqref{ptilde}, \eqref{rhotilde} and \eqref{ntilde} imply that the first law of thermodynamics for an isentropic fluid with a conserved particle number,
\be
d\left(\frac{\tilde \rho}{\tilde n}\right)+\tilde pd\left(\frac{1}{\tilde n}\right)=0\,,
\ee
is verified. Therefore, the pure k-essence Lagrangian $\mathfrak{L}_{\rm m}\left(X\right)$ describes an irrotational perfect fluid with conserved particle number and constant entropy per particle. 

\section{Conclusions}\label{sec:conc}

In this paper we considered theories of gravity in which the first order variation of the trace of the energy-momentum tensor with respect to the metric is a source of the gravitational field. Using $f(R,T)=R+\mathcal F(T)$ gravity minimally coupled to a perfect fluid described by a pure k-essence matter Lagrangian as a well controlled illustrative example, we have shown that the first order variation of the trace of the energy-momentum tensor with respect to the metric depends not only on the proper density, proper pressure and 4-velocity, but also on the sound speed of the fluid. We found that this is so because the second variation of the matter Lagrangian with respect to the metric --- which has been disregarded in several previous studies --- cannot in general be neglected and has a significant dependence on the sound speed. These results are essential for an accurate description of the gravitational and matter fields in the context of $f(R,T)$ gravity --- or of other theories of gravity whose dynamics depends on the first order variation of the trace of the energy-momentum tensor with respect to the metric --- when considering cosmic matter fields described by perfect fluids whose pressure is a function of the proper particle number density alone.

\begin{acknowledgments}

We thank Rui Azevedo, Vasco Ferreira, and my colleagues of the Cosmology group at Instituto de Astrofíısica e Ciências do Espaço for enlightening discussions on modified gravity. We acknowledge the support by Fundação para a Ciência e a Tecnologia (FCT) through the research Grants No. UIDB/04434/2020 and No. UIDP/04434/2020. This work was also supported by FCT through the R$\&$D project 2022.03495.PTDC - {\it Uncovering the nature of cosmic strings}.

\end{acknowledgments}
 
\bibliography{fRTcs}

\begin{thebibliography}{22}
\expandafter\ifx\csname natexlab\endcsname\relax\def\natexlab#1{#1}\fi
\expandafter\ifx\csname bibnamefont\endcsname\relax
  \def\bibnamefont#1{#1}\fi
\expandafter\ifx\csname bibfnamefont\endcsname\relax
  \def\bibfnamefont#1{#1}\fi
\expandafter\ifx\csname citenamefont\endcsname\relax
  \def\citenamefont#1{#1}\fi
\expandafter\ifx\csname url\endcsname\relax
  \def\url#1{\texttt{#1}}\fi
\expandafter\ifx\csname urlprefix\endcsname\relax\def\urlprefix{URL }\fi
\providecommand{\bibinfo}[2]{#2}
\providecommand{\eprint}[2][]{\url{#2}}

\bibitem[{\citenamefont{Aghanim et~al.}(2020)}]{Planck:2018vyg}
\bibinfo{author}{\bibfnamefont{N.}~\bibnamefont{Aghanim}} \bibnamefont{et~al.}
  (\bibinfo{collaboration}{Planck}), \bibinfo{journal}{Astron. Astrophys.}
  \textbf{\bibinfo{volume}{641}}, \bibinfo{pages}{A6} (\bibinfo{year}{2020}),
  \bibinfo{note}{[Erratum: Astron.Astrophys. 652, C4 (2021)]},
  \eprint{1807.06209}.

\bibitem[{\citenamefont{Alam et~al.}(2021)}]{eBOSS:2020yzd}
\bibinfo{author}{\bibfnamefont{S.}~\bibnamefont{Alam}} \bibnamefont{et~al.}
  (\bibinfo{collaboration}{eBOSS}), \bibinfo{journal}{Phys. Rev. D}
  \textbf{\bibinfo{volume}{103}}, \bibinfo{pages}{083533}
  (\bibinfo{year}{2021}), \eprint{2007.08991}.

\bibitem[{\citenamefont{Brout et~al.}(2022)}]{Brout:2022vxf}
\bibinfo{author}{\bibfnamefont{D.}~\bibnamefont{Brout}} \bibnamefont{et~al.},
  \bibinfo{journal}{Astrophys. J.} \textbf{\bibinfo{volume}{938}},
  \bibinfo{pages}{110} (\bibinfo{year}{2022}), \eprint{2202.04077}.

\bibitem[{\citenamefont{Abbott et~al.}(2024)}]{DES:2024tys}
\bibinfo{author}{\bibfnamefont{T.~M.~C.} \bibnamefont{Abbott}}
  \bibnamefont{et~al.} (\bibinfo{collaboration}{DES}) (\bibinfo{year}{2024}),
  \eprint{2401.02929}.

\bibitem[{\citenamefont{Adame et~al.}(2024)}]{DESI:2024mwx}
\bibinfo{author}{\bibfnamefont{A.~G.} \bibnamefont{Adame}} \bibnamefont{et~al.}
  (\bibinfo{collaboration}{DESI}) (\bibinfo{year}{2024}), \eprint{2404.03002}.

\bibitem[{\citenamefont{Akrami et~al.}(2020)}]{Planck:2018jri}
\bibinfo{author}{\bibfnamefont{Y.}~\bibnamefont{Akrami}} \bibnamefont{et~al.}
  (\bibinfo{collaboration}{Planck}), \bibinfo{journal}{Astron. Astrophys.}
  \textbf{\bibinfo{volume}{641}}, \bibinfo{pages}{A10} (\bibinfo{year}{2020}),
  \eprint{1807.06211}.

\bibitem[{\citenamefont{Clifton et~al.}(2012)\citenamefont{Clifton, Ferreira,
  Padilla, and Skordis}}]{Clifton:2011jh}
\bibinfo{author}{\bibfnamefont{T.}~\bibnamefont{Clifton}},
  \bibinfo{author}{\bibfnamefont{P.~G.} \bibnamefont{Ferreira}},
  \bibinfo{author}{\bibfnamefont{A.}~\bibnamefont{Padilla}}, \bibnamefont{and}
  \bibinfo{author}{\bibfnamefont{C.}~\bibnamefont{Skordis}},
  \bibinfo{journal}{Phys. Rept.} \textbf{\bibinfo{volume}{513}},
  \bibinfo{pages}{1} (\bibinfo{year}{2012}), \eprint{1106.2476}.

\bibitem[{\citenamefont{Berti et~al.}(2015)}]{Berti:2015itd}
\bibinfo{author}{\bibfnamefont{E.}~\bibnamefont{Berti}} \bibnamefont{et~al.},
  \bibinfo{journal}{Class. Quant. Grav.} \textbf{\bibinfo{volume}{32}},
  \bibinfo{pages}{243001} (\bibinfo{year}{2015}), \eprint{1501.07274}.

\bibitem[{\citenamefont{Nojiri et~al.}(2017)\citenamefont{Nojiri, Odintsov, and
  Oikonomou}}]{Nojiri:2017ncd}
\bibinfo{author}{\bibfnamefont{S.}~\bibnamefont{Nojiri}},
  \bibinfo{author}{\bibfnamefont{S.~D.} \bibnamefont{Odintsov}},
  \bibnamefont{and} \bibinfo{author}{\bibfnamefont{V.~K.}
  \bibnamefont{Oikonomou}}, \bibinfo{journal}{Phys. Rept.}
  \textbf{\bibinfo{volume}{692}}, \bibinfo{pages}{1} (\bibinfo{year}{2017}),
  \eprint{1705.11098}.

\bibitem[{\citenamefont{Nojiri and Odintsov}(2011)}]{Nojiri:2010wj}
\bibinfo{author}{\bibfnamefont{S.}~\bibnamefont{Nojiri}} \bibnamefont{and}
  \bibinfo{author}{\bibfnamefont{S.~D.} \bibnamefont{Odintsov}},
  \bibinfo{journal}{Phys. Rept.} \textbf{\bibinfo{volume}{505}},
  \bibinfo{pages}{59} (\bibinfo{year}{2011}), \eprint{1011.0544}.

\bibitem[{\citenamefont{Olmo}(2011)}]{Olmo:2011uz}
\bibinfo{author}{\bibfnamefont{G.~J.} \bibnamefont{Olmo}},
  \bibinfo{journal}{Int. J. Mod. Phys. D} \textbf{\bibinfo{volume}{20}},
  \bibinfo{pages}{413} (\bibinfo{year}{2011}), \eprint{1101.3864}.

\bibitem[{\citenamefont{Capozziello and
  De~Laurentis}(2011)}]{Capozziello:2011et}
\bibinfo{author}{\bibfnamefont{S.}~\bibnamefont{Capozziello}} \bibnamefont{and}
  \bibinfo{author}{\bibfnamefont{M.}~\bibnamefont{De~Laurentis}},
  \bibinfo{journal}{Phys. Rept.} \textbf{\bibinfo{volume}{509}},
  \bibinfo{pages}{167} (\bibinfo{year}{2011}), \eprint{1108.6266}.

\bibitem[{\citenamefont{Ludwig et~al.}(2015)\citenamefont{Ludwig, Minazzoli,
  and Capozziello}}]{Ludwig:2015hta}
\bibinfo{author}{\bibfnamefont{H.}~\bibnamefont{Ludwig}},
  \bibinfo{author}{\bibfnamefont{O.}~\bibnamefont{Minazzoli}},
  \bibnamefont{and}
  \bibinfo{author}{\bibfnamefont{S.}~\bibnamefont{Capozziello}},
  \bibinfo{journal}{Phys. Lett. B} \textbf{\bibinfo{volume}{751}},
  \bibinfo{pages}{576} (\bibinfo{year}{2015}), \eprint{1506.03278}.

\bibitem[{\citenamefont{Harko and Lobo}(2010)}]{Harko:2010mv}
\bibinfo{author}{\bibfnamefont{T.}~\bibnamefont{Harko}} \bibnamefont{and}
  \bibinfo{author}{\bibfnamefont{F.~S.~N.} \bibnamefont{Lobo}},
  \bibinfo{journal}{Eur. Phys. J. C} \textbf{\bibinfo{volume}{70}},
  \bibinfo{pages}{373} (\bibinfo{year}{2010}), \eprint{1008.4193}.

\bibitem[{\citenamefont{Harko et~al.}(2011)\citenamefont{Harko, Lobo, Nojiri,
  and Odintsov}}]{Harko:2011kv}
\bibinfo{author}{\bibfnamefont{T.}~\bibnamefont{Harko}},
  \bibinfo{author}{\bibfnamefont{F.~S.~N.} \bibnamefont{Lobo}},
  \bibinfo{author}{\bibfnamefont{S.}~\bibnamefont{Nojiri}}, \bibnamefont{and}
  \bibinfo{author}{\bibfnamefont{S.~D.} \bibnamefont{Odintsov}},
  \bibinfo{journal}{Phys. Rev. D} \textbf{\bibinfo{volume}{84}},
  \bibinfo{pages}{024020} (\bibinfo{year}{2011}), \eprint{1104.2669}.

\bibitem[{\citenamefont{Haghani et~al.}(2013)\citenamefont{Haghani, Harko,
  Lobo, Sepangi, and Shahidi}}]{Haghani:2013oma}
\bibinfo{author}{\bibfnamefont{Z.}~\bibnamefont{Haghani}},
  \bibinfo{author}{\bibfnamefont{T.}~\bibnamefont{Harko}},
  \bibinfo{author}{\bibfnamefont{F.~S.~N.} \bibnamefont{Lobo}},
  \bibinfo{author}{\bibfnamefont{H.~R.} \bibnamefont{Sepangi}},
  \bibnamefont{and} \bibinfo{author}{\bibfnamefont{S.}~\bibnamefont{Shahidi}},
  \bibinfo{journal}{Phys. Rev. D} \textbf{\bibinfo{volume}{88}},
  \bibinfo{pages}{044023} (\bibinfo{year}{2013}), \eprint{1304.5957}.

\bibitem[{\citenamefont{Harko et~al.}(2018)\citenamefont{Harko, Koivisto, Lobo,
  Olmo, and Rubiera-Garcia}}]{Harko:2018gxr}
\bibinfo{author}{\bibfnamefont{T.}~\bibnamefont{Harko}},
  \bibinfo{author}{\bibfnamefont{T.~S.} \bibnamefont{Koivisto}},
  \bibinfo{author}{\bibfnamefont{F.~S.~N.} \bibnamefont{Lobo}},
  \bibinfo{author}{\bibfnamefont{G.~J.} \bibnamefont{Olmo}}, \bibnamefont{and}
  \bibinfo{author}{\bibfnamefont{D.}~\bibnamefont{Rubiera-Garcia}},
  \bibinfo{journal}{Phys. Rev. D} \textbf{\bibinfo{volume}{98}},
  \bibinfo{pages}{084043} (\bibinfo{year}{2018}), \eprint{1806.10437}.

\bibitem[{\citenamefont{Avelino and Azevedo}(2018)}]{Avelino:2018rsb}
\bibinfo{author}{\bibfnamefont{P.~P.} \bibnamefont{Avelino}} \bibnamefont{and}
  \bibinfo{author}{\bibfnamefont{R.~P.~L.} \bibnamefont{Azevedo}},
  \bibinfo{journal}{Phys. Rev. D} \textbf{\bibinfo{volume}{97}},
  \bibinfo{pages}{064018} (\bibinfo{year}{2018}), \eprint{1802.04760}.

\bibitem[{\citenamefont{Azevedo and Avelino}(2018)}]{Azevedo:2018nvi}
\bibinfo{author}{\bibfnamefont{R.~P.~L.} \bibnamefont{Azevedo}}
  \bibnamefont{and} \bibinfo{author}{\bibfnamefont{P.~P.}
  \bibnamefont{Avelino}}, \bibinfo{journal}{Phys. Rev. D}
  \textbf{\bibinfo{volume}{98}}, \bibinfo{pages}{064045}
  (\bibinfo{year}{2018}), \eprint{1807.00798}.

\bibitem[{\citenamefont{Haghani et~al.}(2024)\citenamefont{Haghani, Harko, and
  Shahidi}}]{Haghani:2023uad}
\bibinfo{author}{\bibfnamefont{Z.}~\bibnamefont{Haghani}},
  \bibinfo{author}{\bibfnamefont{T.}~\bibnamefont{Harko}}, \bibnamefont{and}
  \bibinfo{author}{\bibfnamefont{S.}~\bibnamefont{Shahidi}},
  \bibinfo{journal}{Phys. Dark Univ.} \textbf{\bibinfo{volume}{44}},
  \bibinfo{pages}{101448} (\bibinfo{year}{2024}), \eprint{2301.12133}.

\bibitem[{\citenamefont{Gon\c{c}alves et~al.}(2024)\citenamefont{Gon\c{c}alves,
  Rosa, and Lobo}}]{Goncalves:2023umv}
\bibinfo{author}{\bibfnamefont{T.~B.} \bibnamefont{Gon\c{c}alves}},
  \bibinfo{author}{\bibfnamefont{J.~a.~L.} \bibnamefont{Rosa}},
  \bibnamefont{and} \bibinfo{author}{\bibfnamefont{F.~S.~N.}
  \bibnamefont{Lobo}}, \bibinfo{journal}{Phys. Rev. D}
  \textbf{\bibinfo{volume}{109}}, \bibinfo{pages}{084008}
  (\bibinfo{year}{2024}), \eprint{2305.05337}.

\bibitem[{\citenamefont{Ferreira et~al.}(2020)\citenamefont{Ferreira, Avelino,
  and Azevedo}}]{Ferreira:2020fma}
\bibinfo{author}{\bibfnamefont{V.~M.~C.} \bibnamefont{Ferreira}},
  \bibinfo{author}{\bibfnamefont{P.~P.} \bibnamefont{Avelino}},
  \bibnamefont{and} \bibinfo{author}{\bibfnamefont{R.~P.~L.}
  \bibnamefont{Azevedo}}, \bibinfo{journal}{Phys. Rev. D}
  \textbf{\bibinfo{volume}{102}}, \bibinfo{pages}{063525}
  (\bibinfo{year}{2020}), \eprint{2005.07739}.

\end{thebibliography}
 	
 \end{document}